\documentclass[12pt]{article}

\usepackage[margin=2.54cm]{geometry}
\usepackage{amsmath}
\usepackage[compress,numbers,square]{natbib}
\usepackage{hyperref}

\def\kahler{K\"ahler } 
\def\mpl{M_{\mathrm{pl}}}

\def\Q{{\bar Q}}
\def\lsim{\mathrel{\rlap{\lower4pt\hbox{\hskip1pt$\sim$}}
    \raise1pt\hbox{$<$}}}                
\newcommand{\vev}[1]{\langle #1\rangle}

\newcommand{\ek}[2]{e^{{#1}K/{#2}\mpl^2}\,}
\newcommand{\ekp}[2]{e^{{#1}K/{#2}\mpl^2|\phi|^2}\,}

\newcommand{\pdln}[2]{\frac{\partial \ln #1}{\partial \ln #2}}
\newcommand{\pdd}[3]{\frac{\partial^2 #1}{\partial #2 \partial #3}}
\newcommand{\pdlnd}[3]{\frac{\partial^2 \ln #1}{\partial \ln #2
    \partial \ln #3}}
\newcommand{\ompl}{\mathcal{O} \left(\frac{1}{\mpl^2} \right)}

\newcommand{\Z}{\mathcal{Z}}
\newcommand{\phibar}{\bar{\phi}}
\newcommand{\Phibar}{\bar{\Phi}}
\newcommand{\Fphi}{F_\phi}
\newcommand{\Fphibar}{F_{\phibar}}
\def\Wg{\mathcal{W}}

\newcommand{\Secref}[1]{Sec.~\ref{sec:#1}}
\newcommand{\Appref}[1]{App.~\ref{sec:#1}}
\newcommand{\Eqref}[1]{Eq.~(\ref{eq:#1})}

\begin{document}
\begin{titlepage}
\begin{flushright}
\end{flushright}    

\vskip.5cm
\begin{center}
{\huge \bf Anomaly Mediation from Randall-Sundrum to Dine-Seiberg}

\vskip.1cm
\end{center}
\vskip0.2cm 

\begin{center}
{\bf David Sanford and
Yuri Shirman}        
\end{center}                 
\vskip 8pt                   

\begin{center}
{\it
Department of Physics, University of California, Irvine, CA
92697.} \\

{\tt  dsanford@uci.edu, yshirman@uci.edu}
\end{center}

\vglue 0.3truecm

\begin{abstract}
\vskip 3pt \noindent In this paper we reconsider the derivation of
anomaly mediated supersymmetry breaking.  We work in a general
formalism where the $F$-term of the conformal compensator superfield
is arbitrary. This allows for a continuous interpolation between the
original derivation and a more recent Dine-Seiberg derivation of
anomaly mediation. We show that the physical soft parameters are
independent of the compensator $F$-term and results of two formalisms
agree.  Finally, we discuss the role of supersymmetric and
non-supersymmetric thresholds in the effective low energy Lagrangian
as well as the effects of explicit small mass parameters (such as
$\mu$-term) on the superpartner spectrum.
\end{abstract}

\end{titlepage}

\newpage

\section{Introduction}
\label{sec:intro}

We are now at the beginning of the era when Large Hadron Collider
(LHC) at CERN will probe origin of electroweak symmetry breaking and
new physics at the TeV scale.  Among the many proposed scenarios of
interest in coming years, supersymmetry (SUSY) at the TeV scale
represents one of the best motivated extensions of the Standard Model
promising a solution to the gauge hierarchy problem and natural
unification of gauge couplings.  Nevertheless, many supersymmetric
models must be fine-tuned to avoid constrants imposed by the existing
experimental data.  Of these, one of the strongest arises from
stringent bounds on flavor changing neutral currents (FCNCs).  General
SUSY-breaking effects result in FCNCs significantly above current
constraints, so suppression of these effects must be addressed in SUSY
breaking scenarios.  There are several SUSY breaking mechanisms which
naturally suppress FCNCs, including gauge, gaugino, and anomaly
mediated supersymmetry breaking.

Anomaly mediated supersymmetry breaking (AMSB) in particular has been
a subject of active investigation.  The minimal version of AMSB is
very predictive and insensitive to the details of the ultraviolet (UV)
physics, leading to distinct experimental
signals~\cite{Randall:1998uk,Giudice:1998xp}, however it suffers from
negative slepton mass squared problem which is difficult to solve
without reintroducing UV sensitivity
\cite{Pomarol:1999ie,Katz:1999uw,Chacko:1999am,Jack:2000cd,Allanach:2000gu,Chacko:2000wq,Jack:2000nt,Kaplan:2000jz,Anoka:2003kn,Okada:2002mv}.
Finally, the study of formal aspects of AMSB has been motivated by its
somewhat mysterious nature. Indeed, in contrast to gauge or gaugino
mediated scenarios, one cannot easily draw Feynman diagrams that
generate AMSB soft terms in the Lagrangian.  This explains why the
original papers~\cite{Randall:1998uk,Giudice:1998xp} were followed by
a number of attempts to clarify the dynamics underlying the AMSB
mechanism, most notably a detailed study of the origin of gaugino
masses in \cite{Bagger:1999rd}.

More recently a new perspective on AMSB was provided by Dine and
Seiberg~\cite{Dine:2007me}, who clarified several physical aspects of
AMSB.  In particular, they demonstrated that mediation of SUSY
breaking is not a consequence of any anomaly of the theory -- rather
it is the reliance of original {\em derivation} on anomalous
symmetries that led to a name for AMSB.  At the first glance the
Randall-Sundrum (RS) and Dine-Seiberg (DS) formalisms appear to be
quite different.  For example, the RS derivation is performed within
globally supersymmetric effective field theory with SUSY breaking
introduced through a non-supersymmetric regulator.  In contrast, the
origin of SUSY breaking is more intuitive in the DS derivation, with
soft parameters arising entirely due to non-vanishing $F$-terms of
light dynamical fields in the theory.  Moreover, the RS derivation
implicitly neglects all small supersymmetric mass terms in the
Lagrangian since the SUSY breaking effects are introduced at the UV
cutoff; on the other hand, the DS results are most easily obtained in
the Higgs phase and small supersymmetric masses appear to become
important because they affect $F$-terms of the light fields.  In fact,
it has been argued in the literature \cite{deAlwis:2008aq} that the RS
and DS formalisms are not equivalent and at least in some cases lead
to different predictions for soft terms.

In this paper we attempt to better understand the physics underlying
the AMSB mechanism.  To this end we will study the full supergravity
Lagrangian with an arbitrary compensator $F$-term $\Fphi$, which will
allow us to interpolate between the RS and DS formalisms.  We will
show that perturbatively generated soft terms are indeed independent
of $\Fphi$ and identical results are obtained in both the RS and DS
limits.  We will also extend the analysis of the role of
non-supersymmetric thresholds \cite{Pomarol:1999ie,Katz:1999uw} to the
general formalism in the context of a complete SUGRA Lagrangian, and
elaborate on the effects of small supersymmetric mass parameters (such
as a $\mu$-term) on the soft terms in the low energy effective
Lagrangian. In particular, we will show that electroweak scale
$\mu$-term can only result in small corrections to the AMSB
predictions. Even these small corrections are largely accounted for
when a careful renormalization group evolution using MSSM Lagrangian
is performed from the scales above the gravitino mass down to the TeV
scale.

In section \Secref{dsvrs} we will begin by reviewing the full
supergravity Lagrangian in the presence of a conformal compensator
superfield $\Phi$. We will then write down the scalar potential
without integrating out $\Phi$.  In section \Secref{soft} we will
derive soft masses with an arbitrary $\Fphi$ for a toy model with
non-abelian gauge group.  We calculate gaugino masses in
\Secref{gaugino}, and scalar masses in \Secref{scalar}, and show that
the results agree with the original AMSB derivation.  In
\Secref{thresholds} we will review the discussion
\cite{Pomarol:1999ie, Katz:1999uw} of thresholds in the effective
lagrangian and extend it to a formalism with an arbitrary $\Fphi$. In
particular we will study supersymmetric versus non-supersymmetric and
field-dependent versus field-independent thresholds.  We will argue
that small mass parameters, even if they have a supersymmetric origin,
always lead to non-supersymmetric thresholds and their consequences
must be studied in the presence of relevant AMSB soft terms. We will
aply these arguments to explain the effects of the MSSM $\mu$-term
(whether of supersymmetric or non-supersymmetric origin) on the
superpartner spectrum.  In \Secref{conclusion} we present our
concluding remarks.

\section{Conformal Compensator and the Scalar Potential}
\label{sec:dsvrs}

In this section we will briefly review the super-Weyl symmetry of the
supergravity action and, following \cite{Kaplunovsky:1994fg},
introduce the conformal compensator $\Phi = \phi + \Fphi \theta^2$ in
the Lagrangian.  Our goal here is to write the scalar potential in a
way which allows us to interpolate between the RS limit ($\Fphi =
m_{3/2}$) and the DS limit ($\Fphi = 0$).  As a consequence of the
super-Weyl symmetry the scalar potential is invariant under shifts in
$\Fphi$.  However, we will demonstrate that the $F$-terms of matter
fields are {\em not} invariant under shift in $\Fphi$, which allows
for a transition between the physically equivalent RS and DS limits.

The supergravity Lagrangian has the form~\footnote{We use the notation
  of Wess \& Bagger \cite{Wess:1992cp}.}
\begin{equation}
\mathcal{L} = \int d \Theta^2 2 \mathcal{E} \left[\frac{3}{8} \mpl^2
  \left( \bar{\mathcal{D}}^2 - 8 \mathcal{R} \right) |\Phi|^2 \ek{-}{3}
  + \Phi^3 W + \tau \mathcal{W_\alpha W^\alpha} \right] + h.c.
\end{equation}
Here $K$ is the \kahler potential, $W$ is the superpotential,
$\mathcal{W}$ is the supersymmetric field strength for any gauge multiplet
in the theory with coupling $\tau$, $\mathcal{R}$ is the
supersymmetric Ricci scalar, $\mathcal{E}$ is the determinant of the
supersymmetric vierbein, and $\Theta$ is the superspace coordinate.
The superfield $\Phi$ has been introduced into the supergravity
Lagrangian through the \kahler transformation
\begin{equation}
\label{eq:KWPhi}
K \rightarrow K - 3 \mpl^2 \left( \ln \Phi + \ln \Phibar \right)
\qquad W \rightarrow \Phi^3 W \ .
\end{equation}
The Lagrangian does not contain kinetic terms for $\Phi$, thus it is a
purely auxiliary chiral superfield.  The role of $\Phi$ is to formally
restore super-Weyl symmetry of the Lagrangian
\cite{Kaplunovsky:1994fg}, which is broken both by the anomaly in
super-Weyl transformations as well as explicitly by mass terms in the
Lagrangian.  Thus $\Phi$ is usually referred to as a conformal
compensator.  Explicitly, under a super-Weyl transformation
parameterized by an arbitrary chiral superfield $\xi$,
\begin{equation}
\begin{split}
\mathcal{E} & \rightarrow  e^{6\xi} \mathcal{E} \\
\mathcal{E} \left(\bar{\mathcal{D}}^2 - 8 \mathcal{R} \right) &
\rightarrow   \mathcal{E} \left(\bar{\mathcal{D}}^2 - 8 \mathcal{R}
\right) e^{2 \left(\xi + \bar{\xi} \right)} \\
\Phi & \rightarrow  e^{- 2 \xi} \Phi \\
\mathcal{W} & \rightarrow  e^{-3 \xi} \mathcal{W} \\
X & \rightarrow   X\ ,
\end{split}
\end{equation}
where $X$ is an arbitrary chiral superfield.

The breaking of super-Weyl invariance in realistic supergravity
theories can be parameterized by the vacuum expectation value (vev) of
the scalar component of the chiral compensator, $\vev{\phi}$, while
its $\theta^2$ component is usually set to zero. Instead we will keep
the compensator $F$-term, $\Fphi$, as a free parameter to interpolate
between the RS and DS limits~\footnote{We stress that there is no
  physical significance to different values of $\Fphi$.}.

Gaugino masses were obtained in \cite{Randall:1998uk} with
the help of super-Weyl rescaling of chiral superfields in the low
energy (global) SUSY Lagrangian,
\begin{equation}
\label{eq:rescale}
X \Phi \rightarrow X \ .
\end{equation}
We will perform the same rescaling\footnote{Strictly speaking in an
  interacting quantum Lagrangian non-trivial anomalous dimensions of
  the matter fields must be taken in the account when performing this
  rescaling.}  but will work with the full SUGRA Lagrangian.  As a
result of this rescaling, matter superfields transform as $X
\rightarrow e^{- 2 \xi} X$.  While the transformation in
\Eqref{rescale} removes the compensator from all dimension 4 terms in
the Lagrangian, it does not eliminate $\Phi$ completely.  Rather, all
the explicit mass parameters are now accompanied by appropriate
factors of $\Phi$.  For instance, for a superpotential of the form
\begin{equation}
W (X) = X^3 + M_1 X^2 + M_2^2 X \ ,
\end{equation}
this rescaling produces
\begin{equation}
\Phi^3 W (X) \rightarrow W(X, \Phi) = X^3 + \left( M_1 \Phi \right) X^2
+ \left(M_2 \Phi \right)^2 X \ .
\end{equation}
A similar relation holds for the \kahler potential, $|\Phi|^2 K(X,
\bar{X}) \rightarrow K(X, \bar{X}, \Phi, \Phibar)$, with factors of
$\Phi$ and $\Phibar$ appearing alongside mass terms in the \kahler
potential.

As shown in the \Appref{deriv}, the transformation in \Eqref{rescale}
modifies the superspace derivatives according to
\begin{equation}
D_i W  =  W_i + K_i \frac{W}{\mpl^2|\phi|^2} + \ekp{-}{3} \phibar
\Fphibar \partial_{\phibar} \left( \frac{K_i}{\phibar} \right)\,.
\end{equation}
This introduces both explicit and implicit dependence of the scalar
potential on $\Fphi$,
\begin{equation}
 \label{eq:scalarVwphi}
\begin{split}
 V_{\mathrm{scalar}} & =  e\ \ekp{}{3} \left[ K_{ij^*}^{-1}
  D_i W \bar{D}_{j^*} \bar{W} - 3 \frac{|W|^2}{\mpl^2|\phi|^2} \right] \\
 &  + 3 e \mpl^2 |\phi|^2 \ekp{-}{3} |\Fphi|^2 \partial_\phi
\partial_{\phibar} \left[ \ln |\phi|^2 - \frac{K}{3 \mpl^2 |\phi|^2}
  \right] \\
 & - e \left[ \Fphi \partial_\phi W - 3 W \Fphi \partial_\phi
  \left( \ln |\phi|^2 - \frac{K}{3 \mpl^2 |\phi|^2} \right) +
  h.c. \right]\, .
\end{split}
\end{equation}
The implicit dependence on $\Fphi$ enters through the superspace
derivatives $D_iW$, while explicit dependence is present because the
conformal compensator is the only auxiliary field that has not been
integrated out.

Before using \Eqref{scalarVwphi} to interpolate between the RS and DS
limits, we must explicitly break super-Weyl invariance by setting
$\vev{\phi}$=1.  This is necessary to produce a realistic low-energy
Lagrangian, since the physical theory does not exhibit scale
invariance.  The superspace derivative becomes
\begin{equation}
\label{eq:DiW}
D_i W = W_i + K_i \frac{W}{\mpl^2} - \ek{-}{3} \Fphibar K_{ij^*}
X^{*j} = D_i^{\langle 0 \rangle} W - \ek{-}{3} K_{ij^*} X^{*j}
\Fphibar\ ,
\end{equation}
where $D_i^{\langle 0 \rangle} W = W_i + K_i W/\mpl^2$ is the standard
expression for the superspace derivative.  To obtain canonical 
kinetic terms for gravity multiplet, a standard Weyl rescaling must be
performed~\footnote{Since this final rescaling is performed after
  $\vev{\phi}=1$ is set, the fermionic matter fields are rescaled
  accordingly.  However, here we are interested only in scalar fields
  and $F$-terms.}.  After this rescaling, the full scalar potential
takes the form
\begin{eqnarray}
\label{eq:scalarV}
\nonumber V_{\mathrm{scalar}} & = & \ek{}{} \left[ K_{ij^*}^{-1} D_i W
  \bar{D}_{j^*} \bar{W} - 3 \frac{|W|^2}{\mpl^2} \right] - \ek{}{3}
|\Fphi|^2 K_{ij^*} X^i X^{*j} \\
& & + \ek{2}{3} \left[ \Fphi X^i D_i^{\vev{0}} W + h.c. \right]\ ,
\end{eqnarray}
where we have dropped the factor of $e = \det e_m^\mu$ for simplicity
and we used the fact that derivatives of $K$ with respect to $\Phi$
can be expressed in terms of $K_i$, $K_{ij^*}$, etc (see
\Appref{cancellation}).  Moreover, since the rescaled superpotential
is a homogeneous function of chiral superfields (including the
compensator $\Phi$) of order three, the relation
\begin{equation} 
\label{eq:homogeneous}
3 W - X^i W_i - \partial_\phi W = 0
\end{equation}
holds and was used in the last line.  Now a fairly simple calculation
(see \Appref{cancellation}) confirms that the result is independent of
the conformal compensator as expected and reduces to the usual form,
\begin{equation}
 V_{\mathrm{scalar}}^{\langle 0 \rangle} = \ek{}{} \left[
  K^{ij^*} D_i^{\langle 0 \rangle} W \bar{D}_{j^*}^{\langle 0 \rangle}
  \bar{W} - 3 \frac{|W|^2}{\mpl^2} \right]\, .
\end{equation}

Let us restrict our attention to a theory with \kahler potential in
the sequestered form,
\begin{equation}
 \begin{split}
\label{eq:sequestered}
K & =  -3 \mpl^2 |\Phi|^2 \ln \left(1 - \sum\limits_i \frac{|X^i|^2}{3
  \mpl^2 |\Phi|^2} +\frac{f^{\mathrm{hid}}}{3\mpl^2|\Phi^2|}\right) \\
W & =  W_{\mathrm{vis}} + W_{\mathrm{hid}} + W_0 \Phi^3 \ ,
\end{split}
\end{equation}
where $f^{\mathrm{hid}}$ and $W_{\mathrm{hid}}$ are functions of the
hidden sector superfields, $X_i$'s represent visible sector fields,
$W_{\mathrm{vis}}$ is a visible sector superpotential, $W_0$ is the
constant term in the superpotential introduced to cancel the
cosmological constant, and the presence of $\Phi$ is kept explicit.
Supersymmetry is assumed to be broken through the hidden
sector dynamics.  While we stress that the resultant scalar
potential will be independent of the choice of $\Fphi$, we work in the
RS limit to clarify that approach.  This requires that $F$-terms of
all the massless chiral superfields without a vev must be given by the
global SUSY formula, $D_i W = W_i$.  Working with $j$'th superfield we
find that this is achieved by setting
\begin{equation}
\label{eq:fphi}
\Fphibar = \left\langle \ek{}{3} K^{ij^*} K_i
\frac{1}{X^{*j}} \frac{W}{\mpl^2} \right\rangle = \frac{\langle W
  \rangle}{\mpl^2} + \ompl = m_{3/2} + \ompl \,,
\end{equation}
and we stress that there is no summation over $j^*$ in the above
expression. The final result is independent of $j^*$ due to the
assumption that the tree level \kahler potential for all the visible
sector fields is canonical up to Planck suppressed corrections.  As
usual, one must assume that there are no Planck-scale vevs even in the
hidden sector \cite{Bagger:1999rd},
as the presence of such a vev would shift the value of $\Fphi$.

With this choice of $\Fphi$ the scalar potential 
becomes
\begin{equation}
\begin{split}
V_{\mathrm{scalar}} & =  \ek{2}{3} W_i \bar{W}_{i^*} - 3
\ek{}{} \mpl^2 m_{3/2}^2 + m_{3/2}^2 \sum\limits_i |X^i|^2 \\
&  - m_{3/2} \left(\partial_\phi W_{\mathrm{vis}} + \partial_\phi
W_{\mathrm{hid}} + h.c.\right) +
\ompl \, ,
\end{split}
\end{equation}
where \Eqref{homogeneous} was used to simplify the last line.  The
explicit $\mathcal{O}(m_{3/2}^2)$ masses in the third term of the
equation above is cancelled by expanding the exponents in the first
two terms, so any $\mathcal{O}(m_{3/2}^2)$ tree-level contribution to
scalar masses vanishes from the potential.

Finally, we expand the \kahler potential to leading order in
$1/\mpl^2$ and adjust the gravitino mass to cancel the cosmological
constant
\begin{equation}
\label{eq:gravitinomass}
m_{3/2} \equiv \frac{W}{\mpl^2}=\left(\frac{\left\langle K^{ij^*}W_i
  \bar{W}_{i^*} \right\rangle}{3 \mpl^2}\right)^{1/2} + \ompl \, .
\end{equation}

This allows us to write the scalar potential for the visible sector fields
\begin{equation}
 \label{eq:lowenergyscalarV}
V_{\mathrm{vis}}  =  \partial_i W_{\mathrm{vis}} \partial_{i^*}
\bar{W}_{\mathrm{vis}} - m_{3/2} \partial_\phi W_{\mathrm{vis}} -
m_{3/2} \partial_{\phibar} \bar{W}_{\mathrm{vis}} +
\ompl \, .
\end{equation}
We see that at the tree level the full supergravity scalar potential
reduces to the globally supersymmetric potential with addition of a
$m_{3/2}$-dependent holomorphic soft terms.  Note that derivatives
with respect to $\Phi$ vanish for all dimension four terms in the
superpotential. Thus at the tree level all SUSY breaking effects are
associated with the explicit mass parameters in the superpotential in
agreement with \cite{Randall:1998uk}.

\section{Soft mass terms}
\label{sec:soft}

Having written scalar potential with the conformal compensator, we
will now generalize the formalism of \cite{Dine:2007me} to allow for
arbitrary $\Fphi$.  

We will consider a toy model based on an $SU(N)$ gauge group. The DS
formalism in the Higgs phase requires existence of moduli fields whose
vevs provide non-supersymmetric thresholds where soft masses are
generated. Thus we will include $N_f$ vector-like flavors of quarks
$Q$ and $\Q$ in the fundamental representation of the gauge group as
well as a gauge singlet superfield $S$. We will assume that $S$
couples to one of the flavors through the superpotential
\begin{equation}
\label{eq:toyexample}
 W=\lambda_1 S Q_1\bar Q_1 +\frac{M_S}{2}S^2+\frac{\lambda_S}{3} S^3\,.
\end{equation}
For simplicity we will assume that remaining quark flavors ($i =
2,\dots , N_f$) live at the origin of the moduli space.  Depending
on the choice of parameters we now have two types of moduli to
consider:

\begin{itemize}
\item If $\lambda_1=0$, $S$ decouples from the gauge dynamics.  Fields
  $Q_1$ and $\bar Q_1$ are massless and in SUSY limit have an
  arbitrary vev. The gauge group is broken to $SU(N-1)$ and the
  analysis of the soft masses of the light fields is nearly identical
  to that of \cite{Dine:2007me} with a simple generalization to
  non-abelian groups and the addition of $\Fphi$-dependent terms.
\item If $\lambda_1 \neq 0$ the singlet does not decouple. In the
  limit $M_S=\lambda_S=0$ it has an arbitrary vev and provides a
  non-supersymmetric threshold at which quark multiplets become
  heavy. As we shall see this limit provides the generization of the
  DS formalism to a symmetric phase of the gauge theory.
\end{itemize}

\subsection{Gaugino Masses}
\label{sec:gaugino}

Let us start by considering the toy model in a broken phase, $\langle
Q_1 \rangle = \langle \bar Q_1 \rangle=v$.  The gauge
group is broken to an $SU(N-1)$ subgroup but  the 
modulus $Q_1 \bar Q_1$ remains a part of low energy physics and
provides a non-supersymmetric threshold for other light degrees of freedom.
At this threshold  the gauge coupling
constant is
\begin{equation}
\label{eq:tau}
\tau = \frac{1}{g^2} + \frac{b_0}{32\pi^2} \ln \frac{Q_1 \bar
  Q_1}{\Lambda^2 \Phi^2}\ ,
\end{equation}
where $\Lambda$ is the UV cutoff.

Let us justify the inclusion of the conformal compensator $\Phi$ in
\Eqref{tau}.  As has been pointed out in \cite{deAlwis:2008aq}, the
effects of RG evolution between two explicit mass scales in the
Lagrangian are already invariant under super-Weyl transformation. This
is because all mass parameters transform identically. For example, in
our formalism, all the mass terms must come with the appropriate
powers of the conformal compensator but the ratio $M_1/M_2$ is
independent of $\Phi$.  On the other hand, IR cutoffs not associated
with the explicit mass terms in the Lagrangian are qualitatively
different.  Such thresholds must be treated as field dependent and
must be expressed in terms of the light fields. Indeed, this is what
has been done in \Eqref{tau}. The gauge coupling function must be
invariant under the super-Weyl transformation but after the rescaling
performed in \Eqref{rescale} the modulus transforms non-trivially.
Thus super-Weyl invariance requires the inclusion of $\Phi$. Identical
arguments imply that wave-function renormalization factors must depend
on the compensator, $\Z=\Z(|S|/\Lambda|\Phi|)$.

The effective Lagrangian of the theory at the
scale $v$ is given by
\begin{equation}
\label{eq:effW2}
\mathcal{L}_{\mathrm{gauge}} = \int d^2\theta \tau(Q_1, \bar
Q_1,\Lambda) \Wg_\alpha \Wg^\alpha= \int d^2\theta \left(\frac{1}{g^2} +
\frac{b_0}{32\pi^2} \ln \frac{Q_1 \bar
  Q_1}{\Lambda^2\Phi^2}\ \right)\, \Wg_\alpha \Wg^\alpha\,,
\end{equation}
and we stress that $b_0$ is one loop beta-function coefficient of the
theory above $v$, i.e. it includes the contribution of $Q_1$ and $\bar
Q_1$ along with those of the remaining light fields.  Using the
results of the previous section we can obtain a general expression for
the moduli $F$-terms,
\begin{equation}
\begin{array}{lcl}
F_{Q_1}  &=&  - \partial_{Q_1} W - Q_1^* \left( \frac{W}{\mpl^2} - F_{\phi}
\right) + \ompl \\
F_{\Q_1} &=&  - \partial_{\Q_1} W - \Q_1^* \left( \frac{W}{\mpl^2} - F_{\phi}
\right) + \ompl \,.
\end{array}
\end{equation}

Gaugino mass receives the contributions both from the moduli $F$-terms
and factors of $\Fphi$ associated with the UV cutoff $\Lambda$.
Performing the superspace integral in \Eqref{effW2} gives:
\begin{eqnarray}
\label{eq:m32cancellation}
\nonumber m_{\mathrm{gaugino}} & = & \frac{b_0}{32\pi^2}
\left\langle\frac{F_{Q_1}}{Q_1} + \frac{F_{\Q_1}}{\Q_1} \right\rangle -
\frac{b_0 \Fphi}{16\pi^2} \\
 & = & - \frac{b_0}{32\pi^2} \left\langle \frac{2 W}{\mpl^2} +
\frac{\partial_{Q_1} W}{Q_1} + \frac{\partial_{\Q_1} W}{\Q_1} - 2 \Fphi
\right\rangle - \frac{b_0 \Fphi}{16\pi^2} \\
\nonumber & = & - \frac{b_0}{32\pi^2} \left\langle \frac{2 W}{\mpl^2}
+ \frac{\partial_{Q_1} W}{Q_1} + \frac{\partial_{\Q_1} W}{\Q_1} \right\rangle
\simeq - \frac{b_0}{16\pi^2} m_{3/2}\, ,
\end{eqnarray}
where in the second equality we assumed that $\langle Q_1 \rangle$ and
$\langle \Q_1 \rangle$ are real and in the last equality we assume
$\langle \partial_Q W/Q \rangle \ll m_{3/2}$.  Since $\Fphi$ amounts
to a gauge choice soft terms must be independent of it. Indeed we see
that gaugino masses are determined by the SUSY breaking order
parameter, $m_{3/2}$, as expected.  In the DS limit one clearly sees
that gaugino mass is associated with $F$-terms of the dynamical
fields. In the RS limit, the soft mass appears to arise from
non-supersymmetric regulator. Yet the agreement between the two limits
is transparent in our formalism.

We now generalize these results to an unbroken phase of the theory.
Thus instead of breaking the gauge group by the modulus vev at the
non-supersymmetric threshold, we will assume that one quark flavor
becomes heavy due to the superpotential given in
\Eqref{toyexample}. If $M_S = 0$ and $\lambda_S = 0$, the gauge
singlet $S$ is a pseudo-modulus and can acquire large vev~\footnote{We
  will not discuss here dynamics responsible for stabilization of $S$
  at a finite vev.}, $v\gg m_{3/2}$.  The singlet remains a part of
low energy theory even as the quark fields become heavy and decouple
from supersymmetric dynamics. The Lagrangian for the gauge multiplet
becomes
\begin{equation}
\label{eq:effW2S}
\mathcal{L}_{\mathrm{gauge}} = \int d^2\theta \tau(S,\Lambda) \Wg_\alpha
\Wg^\alpha= \int d^2\theta \left(\frac{1}{g^2} + \frac{b_0}{16\pi^2} \ln
\frac{\lambda_S S}{\Lambda \Phi}\ \right)\, \Wg_\alpha
\Wg^\alpha\,,
\end{equation}
and performing the superspace integral we recover AMSB prediction of
gaugino masses which is, once again, independent of the $\Fphi$.

Finally, one can ask how AMSB the prediction of gaugino mass can be
reproduced in a pure super-Yang-Mills theory when neither
(\ref{eq:effW2}) nor (\ref{eq:effW2S}) is directly applicable. As we
have argued, IR thresholds must be treated as fluctuations of light
fields in the low energy theory. In a theory with broken SUSY such
thresholds are accompanied by non-vanishing SUGRA $F$-terms leading to a
correct prediction of gaugino mass.  If no such $F$-terms exist, as in a
pure super-Yang-Mills theory, then one must resort ot explicit
calculation of non-local counterterms.  In fact, in theories where
$F$-terms exist, DS argued that the contribution calculated by using
non-local counterterms is equivalent to that produced by the gauge
coupling function in the Higgs phase \cite{Dine:2007me}, so our
results hold using either method of calculation.

Before concluding this section we would like to point out that
field-dependent thresholds given by the vevs of $Q$, $\bar Q$, and $S$
represent UV cutoffs in the {\em Wilsonian} action for light degrees
of freedom. Thus soft terms generated at these thresholds should be
interpreted as UV effects \cite{Dine:2007me}.

\subsection{Scalar Masses}
\label{sec:scalar}

Soft gaugino masses arise at one loop due to the gauge coupling
renormalization. Thus only the knowledge of the tree level scalar
potential is necessary to obtain the leading contribution to gaugino
masses while the knowledge of wave-function renarmalization is
required to calculate the gaugino masses to higher order in loop
expansion.  As with gaugino masses, the scalar soft terms arise at the
loop level.  However, even the leading, two loop, contribution to
scalar masses requires the knowledge of the wave function
renormalization of the matter fields (albeit at one loop order).

Since the tree level \kahler potential is assumed to be of sequestered
form, tree level contributions vanish, as discussed in \Secref{dsvrs}.
We will calculate perturbative contributions to soft scalar masses in
the unbroken phase of the theory (the generalization to the Higgs
phase is trivial).  Thus we assume that $S$ acquires a vev and the
renormalized \kahler potential at the field-dependent threshold has
the form
\begin{equation}
\label{eq:Krenormalized}
K = - 3 \mpl^2 \ln \left(1 - \frac{|S|^2}{3 \mpl^2 |\Phi|^2} -
\sum\limits_i \frac{|Q_i|^2 + |\bar Q_i|^2}{3 \mpl^2 |\Phi|^2} \Z_i
\left(\frac{|\lambda_S S|}{\Lambda |\Phi|}\right) -
\frac{f^{\mathrm{hid}}}{3\mpl^2 |\Phi|^2} \right)\,
\end{equation}
where $\Z_i$ are wave-function renormalization factors arising due to
RG evolution between UV and IR thresholds, $\Lambda$ and $\lambda_S
\vev{S}$ respectively.  Here we are not interested in the mass of $S$
itself, so to simplify the formulas we have set $\Z_S=1$ at the IR
threshold. As explained earlier, the inclusion of $\Phi$ in $\Z_i$ is
required to maintain super-Weyl invariance.  Furthermore, we can
restrict our attention to the visible sector fields only due to the
assumption of the absence of Planck-scale vevs in the hidden sector.

It is possible to perform the full supergravity calculation of soft
scalar masses for an arbitrary value of $F_\phi$. But since the full
scalar potential is independent of $\Fphi$ we will, in this case, work
in the DS limit, $\Fphi=0$, to simplify the formulas. This limit has
the advantage that the visible sector SUSY breaking is encoded in a
non-zero $F$-term for the modulus $S$~\footnote{This is not generally
  true for an arbitrary $\Fphi$, as factors of $\Fphi$ are present in
  the $F$-terms of other visible sector fields.}. Thus soft scalar
masses arise due to the $F$-terms of the dynamical degrees of freedom.
The expressions for all the required derivatives of the \kahler
potential are presented in the \Appref{kahler}, but the only relevant
term here is the coefficient of $|F_S|^2$,
\begin{equation}
K^{SS^*}=\frac{\ek{-}{3}}{1 + \frac{1}{4} \sum \Z_i
  \dot{\gamma_i} \frac{|Q_i|^2 + |\bar{Q_i}|^2}{|S|^2}} + \ompl \,,
\end{equation}
where we used the standard expressions for anomalous dimensions, 
\begin{equation}
\begin{split} 
\gamma_i & = \pdln{\Z_i}{|S|}= 2\pdln{\Z_i}{S}=2\pdln{\Z_i}{S^*}
\\
\dot{\gamma_i} & = \frac{\partial^2\ln \Z}{\partial |S|^2}=4
\pdlnd{\Z_i}{S}{S^*} = \left[\frac{4}{\Z_i} \pdd{Z_i}{\ln S}{\ln
    S^*} - \gamma_i^2\right] \,,
\end{split}
\end{equation}
Due to the presence of the off-diagonal terms
in the \kahler metric (see \Appref{kahler}), $K^{SS^*}$ does not
depend on anomalous dimensions of the quark superfields, but it does
depend on their derivatives. The scalar potential contains the terms
of the form
\begin{equation}
 V \supset \ek{}{} K^{SS^*} D_S W \bar D_{S^*} \bar W \supset
 -\frac{1}{4} \sum\limits_i \Z_i\dot\gamma_i\left (|Q_i|^2 +|\bar
 Q_i|^2\right) m_{3/2}^2\,.
\end{equation}
This result reproduces famous expression for AMSB soft scalar mass
squareds
\begin{equation}
 \tilde m^2_i = - \frac{1}{4} \dot \gamma m_{3/2}^2\,.
\end{equation}

We can also obtain expressions for AMSB contributions to holomorphic
soft terms. As shown in \Appref{kahler} the following relation holds:
\begin{equation}
\label{eq:kijki}
K^{iS^*} K_{S^*} = (1 - \frac{1}{2} \gamma_i) Q^i\,.
\end{equation}
Substituting this into the formula for scalar potential we have
\begin{equation}
 \begin{split}
  V & \supset  \ek{}{} \left[ K^{ij^*} W_i K_{j^*} \frac{\bar W}{\mpl^2} -
  3 W \frac{\bar W}{\mpl^2} + h.c. \right] \\
& \supset  m_{3/2} \left[ \left( 1 - \frac{1}{2} \gamma_i \right)
  \left( Q^i \partial_{Q^i} W + \Q^i \partial_{\Q^i} W \right) + S W_S
  - 3 W + h.c. \right] \\
& \supset  - m_{3/2} \left[ \partial_\phi W + \sum\limits_i
  \frac{1}{2} \gamma_i \left( Q^i \partial_{Q^i} W + \Q^i
  \partial_{\Q^i} W \right) + h.c. \right]
\end{split}
\end{equation}
The first term in the last expression is already present at tree level
in \Eqref{lowenergyscalarV} while the second term gives one loop
corrections to $A$ and $B$-terms in agreement with
\cite{Randall:1998uk}.

Before concluding this section we note that both gaugino and scalar
masses depend on $\lambda_1$ and $\vev{S}$ only through the anomalous
dimensions of the matter fields.  Thus we can take the limit
$\lambda_1\rightarrow 0$. In this limit $S$ decouples, the quarks
$Q_1$ and $\Q_1$ become light but the prediction for the AMSB mass
terms remains unchanged. This explains why AMSB results still depend
on the contribution of the ``heavy'' fields $Q_1$ and $\Q_1$, in
particular why the gaugino mass is proportional to the
$\beta$-function of the full theory.

\section{Threshold Effects}
\label{sec:thresholds}

In this section we extend discussion of supersymmetric and
non-supersymmetric thresholds found in
\cite{Pomarol:1999ie,Katz:1999uw} to the formulation of AMSB with an
arbitrary $\Fphi$.  Once again we will take SUSY QCD as our toy model.
Let us assume first that one of the quark flavors is given a large
mass term, $M\gg m_{3/2}$. Thus the minimum of the scalar potential is
found at $Q_1=\bar Q_1=0$. For sufficiently large $M$, heavy squarks
obtain non-holomorphic soft mass squareds, $m_{3/2} M$ while their
$F$-terms vanish in both the RS and DS limits. The inclusion of small
soft non-holomorphic masses ($\tilde m \lsim m_{3/2}$) does not shift
the ground state.  Thus threshold associated with the mass of the
heavy quarks is supersymmetric and does not contribute to gaugino
masses; heavy flavor decouples from the low energy physics.

As a next step, consider a model where the quark mass term is given by
a vev of a heavy field. We take the superpotential \Eqref{toyexample}
in the limit of $M_S \gg m_{3/2}$.  At the minimum of the potential
\begin{equation}
\vev{S}= - [ M + m_{3/2} + \mathcal{O}(m_{3/2}^2/M_S) ]
/\lambda_S\,, \hspace*{1cm} \vev{Q_1 \bar Q_1}=0\,,
\end{equation}
and quark superfields become heavy.  While the quark $F$-terms still
vanish in both the RS and DS limits, $F_S$ depends on $\Fphi$.  In
particular, in the DS limit $F_S=0$, thus the threshold at $\vev{S}$
is supersymmetric and does not contribute to soft terms. In the RS
limit, however, $F_S=m_{3/2}\vev{S}$. The supersymmetric nature of the
threshold can be seen instead in the fact that the shift in the ground
state due to SUSY breaking is small~\footnote{This is true in both the
  RS and DS limits.}, $\mathcal{O}(m_{3/2}/M)$. The decoupling of the
supersymmetric threshold is a bit more subtle in the RS limit, where
soft masses receive contributions both from the field-dependent
threshold at $\vev{S}$ and from the regulator. The former is
proportional to $F_S/\vev{S}\approx m_{3/2}$ while the latter is
proportional to $\Fphi=-m_{3/2}$, and we easily see the cancellation
of the two contributions.

We now take a limit $M_S\rightarrow 0$ and $\lambda_S\rightarrow 0$
while keeping $\vev{S}$ fixed.  This is the model of section
\ref{sec:soft} and we already know that soft masses will be
generated. In the DS limit it is easy to see that the origin of the
soft terms lies in the non-supersymmetric nature of the threshold at
$\vev{S}$, since $F_S\approx \vev{S} m_{3/2}$.  On the other hand, in
the RS limit $F_S=0$. Nevertheless, since $S$ is light the location of
the vacuum depends sensitively on the small SUSY breaking terms in the
potential and the threshold is non-supersymmetric. The vanishing of
$F_S$ in the RS limit implies that the contribution of the regulator
can not be canceled by the field-dependent threshold. Thus despite the
apparent generation of AMSB soft terms by the regulator in the RS
limit, the origin lies in the non-supersymmetric nature of an IR
threshold.

We now turn to a generalization of our toy model with several
thresholds. Consider the toy model with the superpotential
\begin{equation}
W = \lambda_1 S_1 Q_1 \Q_1 + \lambda_2 S_2 Q_2 \Q_2 + \lambda_3 S_3
Q_3 \Q_3 + W(S_i) \,.
\end{equation}
We will assume that the gauge singlet fields acquire hierarchical vevs
$S_1<S_2<S_3$ due to $W(S_i)$ so that quark superfields decouple from
the low energy theory and the the threshold at $\vev{S_1}$ is
non-supersymmetric.  Let us consider gaugino masses for concreteness.
The kinetic terms for the gauge multiplet can be written as
\begin{equation}
\label{eq:3thresholds}
 \int d^2\theta \left(\frac{1}{g^2} + \frac{b_0-2}{16\pi^2}
 \ln\frac{\lambda_{1} S_1}{\lambda_{2} S_2} +
 \frac{b_0-1}{16\pi^2} \ln\frac{\lambda_{2} S_2}{\lambda_{3} S_3}
 + \frac{b_0}{16\pi^2} \ln\frac{\lambda_{3} S_3}{\Lambda\Phi}
 \right) \Wg_\alpha \Wg^\alpha\,,
\end{equation}
where $b_0$ is the $\beta$-function coefficient of the theory in the
UV. If thresholds $S_2$ and $S_3$ are supersymmetric, in the DS limit
one finds $F_2 = F_3 = F_\phi = 0$; thus the second and third terms of
the \Eqref{3thresholds} do not contribute to gaugino masses.  It is
worth noting that gaugino masses are proportional to $b_0-2$, the
$\beta$-function of the theory above non-supersymmetric threshold.  In
other words, gaugino masses receive contributions from all the light
fields in the theory as well as one supermultiplet, $Q_1$ and $\Q_1$,
that becomes heavy at a non-supersymmetric threshold.

Let us now assume that the threshold at $S_2$ is also
non-supersymmetric.  In this case $F_1/S_1=F_2/S_2=F_\phi$ and the
contribution of the first term in \Eqref{3thresholds} vanishes.  On
the other hand, the second term in \Eqref{3thresholds} results in
gaugino masses proportional to $b_0-1$.

Finally, assume that $S_1$ and $S_3$ are non-supersymmetric thresholds
while $S_2$ is supersymmetric. We can see that there is a
non-vanishing contribution at each of the thresholds. However, the
combined effect of all the thresholds leads to gaugino mass
proportional to $b_0-1$. We conclude that in all the cases the fields
which become massive at supersymmetric thresholds decouple from the
low energy physics while the fields which become massive at
non-supersymmetric thresholds continue to contribute to soft terms
\cite{Pomarol:1999ie, Katz:1999uw}. We also note that one may
construct more complicated models where heavy fields decouple
partially.

The discussion of the field-dependent as well as non-supersymmetric
thresholds has prepared us for an analysis of the role of small
explicit mass parameters in a supersymmetric Lagrangian, for instance
a $\mu$-term in the MSSM Higgs sector.  In the EWSB vacuum, where both
the $\mu$-term and Higgs vev are non-vanishing, one finds $\partial_H
W\ne 0$.  One then might conclude that the presence of a $\mu$-term in
a supersymmetric Lagrangian may modify AMSB prediciton
\cite{deAlwis:2008aq}.  In particular, in the limit of large $\tan
\beta$ it is possible to have $\langle \partial_{H} W/H \rangle \sim
m_{3/2}$ which naively leads to $\mathcal{O}(1)$ corrections to
gaugino mass. As we will now argue, a more careful analysis shows that
such corrections are subleading. Moreover, they are automatically
taken into account when the renormalization group evolution below the
gravitino mass is performed consistently.

Recall that in the presence of SUSY breaking, the $\mu$-term implies
the existence of the $B$-term $B=\mu m_{3/2}$. For small $\mu\ll
m_{3/2}$ the stable minimum of the potential only exists in the
presence of additional contributions to the soft masses in the Higgs
sector. Thus any threshold associated with the $\mu$-term is
necessarily non-supersymmetric. As explained in this section
corrections to $F_H$ due to non-supersymmetric thresholds are small
compared to $m_{3/2}\vev{H}$ and thus lead to a subleading correction
to gaugino mass.

It is interesting to consider what happens as one increases the
$\mu$-term. When $\mu\sim m_{3/2}$, inclusion of non-holomorphic soft
scalar masses in the analysis leads to an $\mathcal{O}(1)$ shift in
$\vev{H}$ and $F_H$. A simple proportionality relation $F_H\sim
\vev{H} m_{3/2}$ does not hold and a more careful calculation is
required to obtain gaugino mass\footnote{This effect can be seen in
  terms of an effective low energy Lagrangian. Such a Lagrangian
  always contains terms with explicit superspace derivatives and they
  become important precisely when $\mu\sim m_{3/2}$.}.  As $\mu$
becomes large compared to $m_{3/2}$, the vev of $H$ is well
approximated by its supersymmetric value, $H$ decouples from low
energy theory and an AMSB prediction becomes valid again (now with no
contribution from $H$).

This discussion allows us to formulate a prescription for a consistent
calculation of AMSB soft terms. These terms must be evaluated using
AMSB formulas at scales somewhat large compared to gravitiono
mass\footnote{In the absence of explicit $\mathcal{O}(m_{3/2})$ mass
  parameters in the visible sector, this calculation can be performed
  at the scale of gravitino mass.}. Values of soft parameters obtained
in such a way should then be treated as a boundary conditions and
detailed RGE calculations with full non-supersymmetric MSSM Lagrangian
must be used to obtain low energy predictions.  The phenomenological
consequences of the $\mu$-term are automatically accounted for In this
approach.

\section{Conclusion}
\label{sec:conclusion}

In this paper we have reviewed anomaly mediated supersymmetry
breaking. We have uplifted the original formalism of
\cite{Randall:1998uk} to the full supergravity Lagrangian and
generalized it to allow an arbitrary compensator $F$-term. This
allowed us to interpolate between the RS and DS derivations of anomaly
mediation and show that they lead to completely equivalent results. We
have also discussed the effects of supersymmetric and
non-supersymmetric thresholds in the theory as well as the role of
small supersymmetric mass parameters in the visible sector of the
theory.

\section*{Acknowledgements}

We thank Arvind Rajaraman for useful discussions. This work was
supported in part by NSF grants PHY-0653656 and PHY-0970173.


\appendix

\section{Conformal Compensator and SUGRA Lagrangian}
\label{sec:deriv}

In this appendix we derive the scalar potential in the supergravity
formalism with a conformal compensator.  This discussion closely
follows \cite{Wess:1992cp}.  The supergravity Lagrangian is
\begin{equation}
\mathcal{L} = \int d \Theta^2 2 \mathcal{E} \left[\frac{3}{8} \mpl^2
  \left( \bar{\mathcal{D}}^2 - 8 \mathcal{R} \right) |\Phi|^2
  \ek{-}{3} + \Phi^3 W + \tau \mathcal{W W} \right] + h.c.
\end{equation}
where $\Phi$ is the conformal compensator which has been introduced
through the replacement
\begin{equation}
K \rightarrow K - 3 \mpl^2 \left( \ln \Phi + \ln \Phibar \right)
\qquad W \rightarrow \Phi^3 W \ .
\end{equation}

Following \cite{Randall:1998uk} we further rescale the chiral
superfields $\Phi X^i \rightarrow X^i$, which causes the matter fields
to transform under super-Weyl transformations.  As a result the
explicit factors of $\Phi$ only appear in association with mass terms
in the Lagrangian.

Super-Weyl transformations consist of
\begin{eqnarray}
\mathcal{E} & \rightarrow & e^{6 \xi} \mathcal{E} \\
\Phi & \rightarrow & e^{- 2 \xi} \Phi \\
X^i & \rightarrow & e^{- 2 \xi} X^i
\end{eqnarray}
and the Lagrangian becomes
\begin{equation}
\mathcal{L} = \int d \Theta^2 2 \mathcal{E} \left[\frac{3}{8} \mpl^2
  \left( \bar{\mathcal{D}}^2 - 8 \mathcal{R} \right) |\Phi|^2 e^{-
    K(X_i,\bar{X_i},\Phi,\Phibar)/3 \mpl^2|\Phi|^2} + W
(X,\Phi) \right] + h.c.\,,
\end{equation}

In this basis the scalar Lagrangian is
\begin{eqnarray}
\nonumber \mathcal{L}_{\mathrm{scalar}} & = & \frac{1}{9} e \Omega
\left|M^* - 3 \left( \ln \Omega \right)_i F_i \right|^2 + e \Omega
\left( \ln \Omega \right)_{ij^*} F_i F_j^* \\
& & + e \left(W_i F_i - W M^* + h.c. \right) \\
\nonumber & = & \frac{1}{9} e \Omega \left|M^* - 3 \left( \ln \Omega
\right)_i F_i - 3 F_{\phi} \partial_\phi \left( \ln \Omega \right)
\right|^2 \\
\nonumber & & + e \Omega \left( \ln \Omega \right)_{ij^*} F_i F_j^* +
e \Omega \partial_{\phibar} \left( \ln \Omega \right)_{i} F_i
\Fphibar \\
\nonumber & & + e \Omega \partial_\phi \left( \ln \Omega
\right)_{j^*} F_j^* \Fphi + e \Omega |\Fphi|^2 \partial_\phi
\partial_{\phibar} \left( \ln \Omega \right) \\
& & + e \left(W_i F_i + \Fphi \partial_\phi W - W M^* +
h.c. \right)\, ,
\end{eqnarray}
where 
\begin{equation}
\label{eq:omega}
\Omega = - 3 \mpl^2
|\Phi|^2 \exp\left({-K(X_i,\bar{X_i},\Phi,\Phibar)/3\mpl^2 |\Phi|^2}\right)\,. 
\end{equation}
and $M$ is an auxiliary scalar field in the gravity multiplet.
Integrating out the combination $N^* = M^* - 3 \left( \ln \Omega
\right)_i F_i - 3 \Fphi \partial_\phi \left( \ln \Omega \right)$
allows us to write the scalar Lagrangian in the form
\begin{equation}
\begin{split}
 \mathcal{L}_{\mathrm{scalar}}  = & - 9 e |W|^2 \Omega^{-1}
- e \Omega^{-1} \left( \ln \Omega \right)_{ij^*}^{-1} D_i W \bar{D}_{j^*}
\bar{W} \\
  & + e \Omega |\Fphi|^2 \partial_\phi
\partial_{\phibar} \left( \ln \Omega \right) + e \Fphi
\partial_\phi W + e \Fphibar \partial_{\phibar} \bar{W} \\
 & - 3 e W \Fphi \partial_\phi \left( \ln \Omega \right) - 3 e W
\Fphibar \partial_{\phibar} \left( \ln \Omega \right)\,,
\end{split}
\end{equation}
where the covariant superspace derivative is defined by
\begin{equation}
D_{j^*} \bar{W}  =  \bar{W}_{j^*} + \Omega \Fphi \partial_\phi
\left( \ln \Omega \right)_{j^*} - 3 W \left( \ln \Omega
\right)_{j^*} \,.
\end{equation}

Finally, using \Eqref{omega} we can write
\begin{eqnarray}
\begin{split} 
 \mathcal{L}_{\mathrm{scalar}}  = & e\, \ekp{}{3} \left[ 3
  \frac{|W|^2}{\mpl^2|\phi|^2} - K_{ij^*}^{-1} D_i W \bar{D}_{j^*}
  \bar{W} \right] \\
 &  - 3 e \mpl^2 |\phi|^2 \ekp{-}{3} |\Fphi|^2 \partial_\phi
\partial_{\phibar} \left[ \ln |\phi|^2 - \frac{K}{3 \mpl^2 |\phi|^2}
  \right] \\
&  + e \left[ \Fphi \partial_\phi W - 3 W \Fphi \partial_\phi \left(
  \ln |\phi|^2 - \frac{K}{3 \mpl^2 |\phi|^2} \right) + h.c. \right] \,
,
\end{split}
\end{eqnarray}
where 
\begin{equation}
D_i W  =  W_i + K_i \frac{W}{\mpl^2|\phi|^2} + \ek{-}{3} \phibar
\Fphibar \partial_{\phibar} \left( \frac{K_i}{\phibar} \right)
\end{equation}

In order to get a canonical gravity kinetic term, the vierbein must be
rescaled as $e_m^a \rightarrow e_m^a \ek{-}{6}$.  However, this
tranformation cannot be performed in an explicitly supersymmetric
manner in the conformal compensator formalism.  We therefore set
$\vev{\phi} = 1$ and then perform a non-supersymmetric Weyl
transformation to achieve canonical gravitational kinetic terms.  This
retains $\Fphi$ as an arbitrary parameter.  We stress that though
doing so breaks the super-Weyl invariance, that invariance has already
been broken explicitly by giving a non-zero value to the background
scalar field $\phi$.

The superspace derivative and scalar potential can now be written as
\begin{equation}
\begin{split}
\label{eq:Scalarwphi}
D_i W = &\,  W_i + K_i \frac{W}{\mpl^2} - \ek{}{3} \Fphibar \left(K_i -
\partial_{\phibar} K_i |_{\phi=1} \right) \\
 V_{\mathrm{scalar}} = &\,  e \,\ek{}{} \left[ K_{ij^*}^{-1} D_i
  W \bar{D}_{j^*} \bar{W} - 3 \frac{|W|^2}{\mpl^2} \right] \\
 &  - e\, \ek{}{3} |\Fphi|^2 \left[\partial_\phi
  \partial_{\phibar} K - \partial_\phi K - \partial_{\phibar} K +
  K\right]_{\phi = 1} \\
 &  - e\, \ek{2}{3} \left[ \Fphi \partial_\phi W - 3 W \Fphi \left(1 +
  \frac{K}{3 \mpl^2} - \frac{\partial_\phi K |_{\phi=1}}{3 \mpl^2}
  \right) + h.c. \right]\, .
\end{split}
\end{equation}

\section{Physical Scalar Potential and Cancellation of $\Fphi$}
\label{sec:cancellation}

In this appendix we  show explicitly that the scalar potential is
independent of $\Fphi$.  Consider an arbitrary \kahler potential of
the form
\begin{equation}
 K = - 3 \mpl^2 |\Phi|^2 \ln \Sigma \, .
\end{equation}
Though any \kahler potential may be expressed in this form, the
function $\Sigma$ has the simplest form when the \kahler potential has
a sequesterd form.  $\Sigma$ is a function of fields and coupling of
the theory and at the same time has mass dimension zero. On the other
hand, in the RS formalism chiral superfields have non-trivial Weyl
weights. Thus to ensure proper transformations of the \kahler
potential under super-Weyl transformations $\Sigma$ must take the form
\begin{equation}
\label{eq:sigmaform}
\Sigma \left( X^i, X^{i*}, \Phi, \Phibar, \{M \} \right) = \Sigma
\left( \frac{X^i}{M\Phi}, \frac{X^{i*}}{M\Phibar} \right) \, ,
\end{equation}
where $X^i$ are the matter fields in both the visible and hidden
sectors and $M$ represents any appropriate mass parameter in the
theory.

We can use \Eqref{sigmaform} to relate the derivatives with respect to
$\phi$ to derivatives with respect to matter fields,
\begin{equation}
\begin{split}
\partial_\phi \Sigma & =  \frac{\partial \Sigma}{\partial \left(X^i /
  \phi\right)} \frac{\partial \left(X^i / \phi\right)}{\phi} = -
\frac{\partial \Sigma}{\partial \left(X^i / \phi\right)}
\frac{X^i}{\phi^2} = - \Sigma_i \frac{X^i}{\phi} \\
\partial_\phi \partial_{\phibar} \Sigma & =  \frac{\partial^2
  \Sigma}{\partial \left(X^i / \phi\right) \partial \left(X^{*j} /
\phibar \right)} \frac{X^i X^{*j}}{|\phi|^4} = \Sigma_{ij^*}
  \frac{X^i X^{*j}}{|\phi|^2} \\
\partial_{\phibar} \Sigma_i & =  - \Sigma_{ij^*}
\frac{X^{*j}}{\phibar} \, .
\end{split}
\end{equation}
These equations then allow us to simplify several of the terms in
\Eqref{Scalarwphi},
\begin{equation}
 \begin{split}
K_i & =  - 3 \mpl^2 |\phi|^2 \frac{\Sigma_i}{\Sigma} \\
K_{ij^*} & =  - 3 \mpl^2 |\phi|^2 \left[ \frac{\Sigma_{ij^*}}{\Sigma}
    - \frac{\Sigma_i \Sigma_{j^*}}{\Sigma^2} \right] \\
K^{ij^*} & =  -\frac{\Sigma}{3 \mpl^2 |\phi|^2} \left[\Sigma_{ij^*} -
  \frac{\Sigma_i \Sigma_{j^*}}{\Sigma} \right]^{-1} \\
\partial_\phi K & =  - 3 \mpl^2 \phibar \left[ \ln \Sigma + \phi
  \frac{\partial_\phi \Sigma}{\Sigma} \right] = \frac{1}{\phi} \left[K
  - X^i K_i \right] \\
\partial_\phi \partial_{\phibar} K & =  - 3 \mpl^2 \left[ \ln
  \Sigma + \phi \frac{\partial_\phi \Sigma}{\Sigma} + \phibar
  \frac{\partial_{\phibar} \Sigma}{\Sigma} + |\phi|^2
  \frac{\partial_\phi \partial_{\phibar} \Sigma}{\Sigma} - |\phi|^2
  \frac{\partial_\phi \Sigma \partial_{\phibar} \Sigma}{\Sigma^2}
  \right] \\
& =  \frac{1}{|\phi|^2} \left[ K - X^i K_i - X^{i*} K_{i^*} +
  K_{ij^*} X^i X^{*j} \right] \\
\partial_{\phibar} K_i & =  - 3 \mpl^2 \phi
\left[\frac{\Sigma_i}{\Sigma} + \phibar \frac{\partial_{\phibar}
    \Sigma_i}{\Sigma} - \phibar \frac{\Sigma_i
    \partial_{\phibar} \Sigma }{\Sigma^2} \right] = \frac{1}{\phibar}
\left[K_i - K_{ij^*} X^{*j} \right] \, .
\end{split}
\end{equation}

Then, setting $\vev{\phi} = 1$, the scalar potential reduces to
\begin{equation}
 \begin{split}
D_i W  = &\, W_i + K_i \frac{W}{\mpl^2} - \Sigma \Fphibar
K_{ij^*} X^{*j} = D_i^{\langle 0 \rangle} W - \Sigma K_{ij^*} X^{*j}
\Fphibar \\
 V_{\mathrm{scalar}} = & \,  \frac{1}{\Sigma^3} \left[
  K_{ij^*}^{-1} D_i W \bar{D}_{j^*} \bar{W} - 3 \frac{|W|^2}{\mpl^2}
  \right] \\
 & - \frac{1}{\Sigma} |\Fphi|^2 K_{ij^*} X^i X^{*j} -
\frac{1}{\Sigma^2} \left[ \Fphi \partial_\phi W - 3 W \Fphi - \Fphi
  \frac{W}{\mpl^2} K_i X^i + h.c. \right] \ .
\end{split}
\end{equation}

We can further simplify the second line of the potential by noting
that, with the inclusion of $\Phi$, $W$ is a homogeneous polynomial
function of order three in the fields.  Thus, $3 W - X^i W_i -
\partial_\phi W = 0$, which gives
\begin{equation}
\begin{split}
 V_{\mathrm{scalar}} = &\, \frac{1}{\Sigma^3} \left[
  K_{ij^*}^{-1} D_i W \bar{D}_{j^*} \bar{W} - 3 \frac{|W|^2}{\mpl^2}
  \right] \\
& - \frac{1}{\Sigma} |\Fphi|^2 K_{ij^*} X^i X^{*j} +
 \frac{1}{\Sigma^2} \left[ \Fphi X^i D_i^{\vev{0}} W + h.c. \right]
 \ .
\end{split}
\end{equation}
This, with the replacement $\Sigma = \ek{-}{3}$, leads to
\Eqref{scalarV}.  It is now fairly simple to show that $V$
is independent of $\Fphi$:
\begin{equation}
 \begin{split}
 V_{\mathrm{scalar}}  = &\, V_{\mathrm{scalar}}^{\langle 0
  \rangle} - \frac{1}{\Sigma^2} \left[X^i K_{ik^*} K^{jk^*} \Fphi
  D_j^{\vev{0}} W + h.c. \right] + \frac{1}{\Sigma} |\Fphi|^2 K_{ik^*}
K^{lk^*} K_{lj^*} X^i X^{*j} \\
 & - \frac{1}{\Sigma} |\Fphi|^2 K_{ij^*} X^i X^{*j} +
\frac{1}{\Sigma^2} \left[ \Fphi X^i D_i^{\vev{0}} W + h.c. \right] \\
 = &\,  V_{\mathrm{scalar}}^{\langle 0 \rangle} \, .
\end{split}
\end{equation}

\section{Renormalized \kahler metric}
\label{sec:kahler}

Here we present the \kahler potential derivatives used in
\Secref{scalar}.  The first and second derivatives are
\begin{align}
K_i^{\langle\mathrm{vis}\rangle} & = \ek{}{3} \Z_i Q_i^* ~~~~~~~~
K_S^{\vev{\mathrm{vis}}}=\ek{}{3}\left(S^* + \frac{1}{2} \sum\limits_i
\frac{|Q_i|^2 + |\bar{Q}_i|^2}{S} \Z_i \gamma_i\right)\nonumber \\
K_{Sj^*}^{\langle\mathrm{vis}\rangle}&= \ek{}{3} \frac{1}{2} \frac{Q_j}{S}
\Z_i \gamma_j ~~~~~~~~ K_{ij^*}^{\langle\mathrm{vis}\rangle}=\ek{}{3} \Z_i
\delta_{ij^*}\\
K_{SS^*}^{\langle\mathrm{vis}\rangle}&= \ek{}{3}\left(1 +
\sum\limits_i \frac{|Q_i|^2 + |\bar{Q}_i|^2}{|S|^2} \pdd{\Z_i}{\ln
  S}{\ln S^*}\right)\,, \nonumber 
\end{align}
where $i$ and $j^*$ correspond to derivatives with respect to all
matter fields besides the modulus $S$.  Note that the anomalous
dimension only appears in $S$-dependent derivatives, since the
wave-function renormalization is dependent only on $S$.  Inverting
$K_{ij^*}$ gives
\begin{equation}
\begin{split}
K^{\vev{\mathrm{vis}}SS^*} & = \frac{\ek{-}{3}}{1 + \frac{1}{4}
  \sum\limits_i \Z_i \dot{\gamma_i} \frac{|Q_i|^2 +
    |\bar{Q_i}|^2}{|S|^2}} + \ompl \\
K^{\langle\mathrm{vis}\rangle iS^*} & = - \frac{\ek{-}{3}}{1 + \frac{1}{4}
  \sum\limits_j \Z_j \dot{\gamma_j} \frac{|Q_j|^2 + |\bar{Q}_j|^2}{|S|^2}}
\frac{Q_i\gamma_i}{2S} + \ompl \\
K^{\langle\mathrm{vis}\rangle ij^*} & =  \frac{\ek{-}{3}}{1 + \frac{1}{4}
  \sum\limits_k \Z_k \dot{\gamma} \frac{|Q_k|^2 +
    |\bar{Q}_k|^2}{|S|^2}} \frac{\delta^{i^*j}}{\Z_i} + \mathcal{O}
\left( \frac{|Q|^2}{|S|^2} \right) + \ompl \,.
\end{split}
\end{equation}
From this, it is apparent that in the DS limit the only relevant term
is $K^{SS^*}$.  Terms involving $K^{ij^*}$ appear to contribute both
at tree level and in perturbation theory. However, tree level
contributions of these terms vanish cancel due to the sequestered form
of the \kahler potential while loop contributions vanish because all
matter fields except $S$ are assumed to have no vevs. In models where
several light fields have vevs, one must consider interplay of
non-supersymmetric thresholds as discussed in section
\ref{sec:thresholds}.

Finally, to derive holomorphic soft terms we need the following
expression:
\begin{equation}
K^{iS^*} K_{S^*} = \left( 1 - \frac{1}{2} \gamma_i \right) Q^i + \mathcal{O}
\left( \frac{|Q|^2}{|S|^2} \right) + \ompl \,.
\end{equation}
Once again we have dropped higher-dimensional terms that do not
contribute to soft terms.


\bibliography{amsbbib}{}

\end{document}